\documentclass[a4paper,12pt]{article}

\pdfoutput=1

\usepackage{amsthm,amsmath}
\usepackage[round]{natbib}
\usepackage[table]{xcolor}  
\usepackage[colorlinks,citecolor=blue,urlcolor=blue]{hyperref}
\usepackage{graphicx}
\usepackage{amsfonts}
\usepackage{amsbsy}
\usepackage{mathrsfs}
\usepackage{lscape}
\usepackage{float}
\usepackage{multirow}

\usepackage{enumerate}
\usepackage{pdfpages}
\usepackage{authblk}
\usepackage{xurl}
\usepackage{footmisc}

\title{Digit analysis for Covid-19 reported data}

\author{Jean-Fran\c cois Coeurjolly}
\affil{Universit\'e du Qu\'ebec \`a Montr\'eal (UQAM), Canada}

\newcommand{\study}{\today}%April, 22nd 2020}
\newcommand{\BB}{5000}

\begin{document}

\maketitle

\begin{quotation}
\noindent {\it Abstract:} The coronavirus which appeared in December 2019 in Wuhan has spread out worldwide and caused the death of more than 280,000 people (as of \study). Since February 2020, doubts were raised about the numbers of confirmed cases and deaths reported by the Chinese government. In this paper, we examine data available from China at the city and provincial levels and we compare them with Canadian provincial data, US state data and French regional data. We consider cumulative and daily numbers of confirmed cases and deaths and examine these numbers through the lens of their first two digits and in particular we measure departures of these first two digits to the Newcomb-Benford distribution, often used to detect frauds. Our finding is that there is no evidence that cumulative and daily numbers of confirmed cases and deaths for all these countries have different first  or second digit distributions. We also show that the Newcomb-Benford distribution cannot be rejected for these data.

\vspace{9pt}
\par

\bigskip

\centerline{ \it Dedicated to people fighting against Covid-19.}

\end{quotation}\par

%%%%%%%%%%%%%%%%%%%%%%%%%%%%%%%%%%%%%%%%%%%%%%%%%%%%%%%%%%%%%%%%%%%%%%%%%%%%%%%%%%%%%
%%%%%%%%%%%%%%%%%%%%%%%%%   Main text     %%%%%%%%%%%%%%%%%%%%%%%%%%%%%%%%%%%%%%%%%%%
%%%%%%%%%%%%%%%%%%%%%%%%%%%%%%%%%%%%%%%%%%%%%%%%%%%%%%%%%%%%%%%%%%%%%%%%%%%%%%%%%%%%%

\section{Introduction} \label{sec:intro}

Coronavirus   disease 2019 (Covid-19) caused by SARS-CoV-2, is an infectious disease that was first identified in December 2019 in Wuhan in the Hubei province of China. The World Health Organisation (WHO) declared the Covid-19 outbreak a Public Health Emergency of International Concern on 30 January 2020 and a pandemic on 11 March 2020.
As reported by the Chinese government on 29 February 2020 the cumulative number of confirmed cases was 79,968 while the cumulative number of deaths was 2,873.\footnote{source: European Centre for Disease Prevention and Control, see Section~\ref{sec:data}.} In the meantime, data reported by the Chinese government have been questioned and suspicion has been raised about the government intentionally hiding the real situation. This idea spread out worldwide in February and March 2020, see for instance articles in \texttt{Radio-Canada}\footnote{\url{https://ici.radio-canada.ca/nouvelle/1690496/Covid-19-chine-wuhan-donnees-verification-decrypteurs}}, \texttt{Le Devoir}\footnote{\url{https://www.ledevoir.com/societe/575070/un-scenario-pire-que-celui-de-la-chine}}, \texttt{New York Times}\footnote{\url{https://www.nytimes.com/2020/04/02/us/politics/cia-coronavirus-china.html}}, \texttt{Bloomberg}\footnote{\url{https://www.bloomberg.com/news/articles/2020-04-01/china-concealed-extent-of-virus-outbreak-u-s-intelligence-says}}, \texttt{Le Monde}\footnote{\url{https://www.lemonde.fr/international/article/2020/03/30/coronavirus-doutes-sur-l-estimation-du-nombre-de-deces-en-chine_6034871_3210.html}} or the wikipedia page about the coronavirus pandemic in mainland China\footnote{\url{https://en.wikipedia.org/wiki/2019-20_coronavirus_pandemic_in_mainland_China}}. A similar question could also be asked for data reported by other countries.

The present time is definitely the era of data scientists and a considerable effort has been made to make data available \citep{alamo2020open} at different levels (national, provincial, etc). The objective of this paper is to provide an empirical comparison for several countries (China, Canada, US and France) and see if we can detect anomalies in such data using first or second digit analysis from these numbers. We would like to emphasize that the intention of this paper is not to single out any country (actually the main conclusion from this paper is that digit distributions from these data look quite similar). In the same way, we do not provide any direct conclusion that there were no frauds in reporting data. What guided the curiosity of the author was to investigate the possible use (or not) of a statistical distribution, namely the Newcomb-Benford distribution, to model digits from Covid-19 data.

\begin{figure}[H]
\begin{center}
\includegraphics[scale=1]{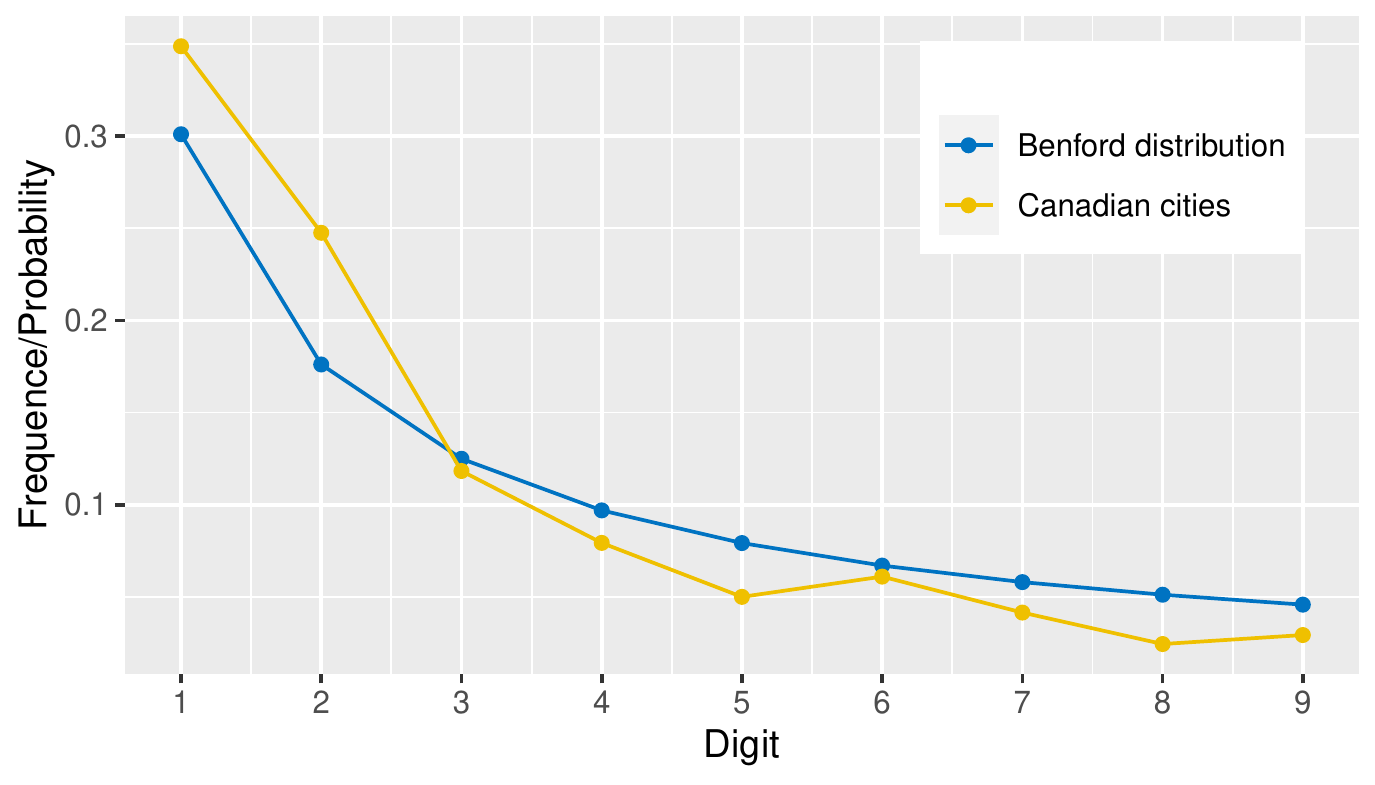}	
\caption{Newcomb-Benford distribution and frequencies for the first digits of population sizes of the 800 largest Canadian cities in 2011.}
\label{cities}
\end{center}
\end{figure}

Pick any series of numbers and make a table of the leading digit (1 for 1234, 4 for 432, etc.) then there is a ``high chance'' that the most frequent digit is 1, then 2, etc. To illustrate this, Figure~\ref{cities} reports frequencies of the leading digit for population sizes of the 800  Canadian largest cities in 2011\footnote{\url{https://www12.statcan.gc.ca/census-recensement/2011/dp-pd/hlt-fst/pd-pl/Table-Tableau}}. Such a surprising phenomenon turns out to be observable in many real-life datasets. 

In 1881, \citet{newcomb1881note} wrote ``That the ten digits do not occur with equal frequency must be evident to any one making much use of logarithmic tables, and noticing how much faster the first pages wear out than the last ones''. \citet{benford1938law} formalized this observation and formulated a distribution for the first leading digit. This distribution, known as the first digit law, is now often referred to as the Newcomb-Benford distribution. Since then, this distribution has been extended in many directions: generalization of this law to numbers expressed in other bases, and also a generalization from leading 1 digit to leading $m$ digits (and actually the joint distribution of the first $m$ digits). In particular Table~\ref{tab:NB} reports the marginal distribution for the first and second digits. Figure~\ref{cities} depicts the distribution of the Newcomb-Benford for the first digit and it can be observed that first digits from Canadian cities sizes in 2011 are not far from a Newcomb-Benford distribution.

\begin{table}[H]
\begin{center}
{\scriptsize\begin{tabular}{rcccccccccc}
\hline
\rowcolor{gray!20}& \multicolumn{10}{c}{Digit} \\
 \rowcolor{gray!20}$k$&0&1&2&3&4&5&6&7&8&9 \\
\hline
$\mathrm P_1(k)$ & & 30.1\% &17.6\% &12.5 \% & 9.7\% &  7.9 \% & 6.7 \% & 5.8 \% & 5.1 \% & 4.6\%\\
$\mathrm P_2(k)$ & 12.0\%& 11.4\%& 10.9 \%&10.4 \%&10.0\%&  9.7\%&  9.3 \%& 9.0 \%& 8.8 \%& 8.5\%\\
\hline
\end{tabular}}
\caption{
Marginal Newcomb-Benford probabilities, denoted by $P_1(\cdot)$ and $P_2(\cdot)$ for the first and second digits.} \label{tab:NB}
\end{center}
\end{table}

An important mathematical and statistical literature, from which \citet{genest2011loi,berger2015introduction,varian1972benfords} constitute excellent overviews, reveals explanations about this phenomenon. An intuition \citep[see e.g.][]{gauvrit2008pourquoi} is that fractional parts of logarithms of numbers tend to be uniformly distributed. Now think reversely: if the fractional part of $\log_{10}(x)$ is uniformly distributed then naturally $x$ has more chance to occur between 1 and 2 than between 9 and 10 and the Newcomb-Benford distribution can be derived in this way.

The Newcomb-Benford distribution has interesting properties: scale invariance \citep{pinkham1961distribution}, connections with mixture of uniform distributions \citep{janvresse2004uniform}, etc. In statistics, \citet{formann2010newcomb} investigates how common distributions are related to the Newcomb-Benford one. Also, like the Gaussian curve for the empirical mean, or the Gumbel distribution for the maximum of random variables, the Newcomb-Benford distribution appears as a natural limit, see for instance \citet{genest2011loi} or \citet{chenavier2018products} (and references therein) for recent developments on this subject.

From a practical point of view, the Newcomb-Benford distribution (and in particular the first digit law) has been used in an attempt to detect frauds in reported numbers. For instance, \citet{deckert2011benford} used it to detect frauds in elections, \citet{el2005price} investigated consumer price digits before and after the euro introduction for price adjustments, \citet{muller2011greece} found out possible frauds in the macroeconomic data the Greek government reported, \cite{diekmann2007not} or \citet{gauvrit2008pourquoi} made use of this distribution to  detect frauds in scientific papers, etc.

In this paper, we investigate the use of the first  and second digit distribution to detect potential anomalies in reported Covid-19 data. Because most of daily and cumulative data are quite large, an analysis of the first two digits separately seems relevant. Analyzing the third digit or analyzing the joint distribution of the first two digits however would drastically reduce the sample size of data and has not been considered. 
We investigate the numbers of confirmed cases and  deaths reported by several countries, more specifically China, Canada, US and France, and aim at detecting potential differences between these countries. Such data are recorded at a national level.  To the best of our knowledge, there is no theory that the first or second digit of epidemiological data should obey the Newcomb-Benford distribution but given the important literature on the Benford's phenomenon, investigating such a question is worthwile. The rest of the paper is organized as follows. The data acquisition and preprocessing is described in Section~\ref{sec:data}. Results are presented in Section~\ref{sec:results} and discussed in Section~\ref{sec:discussion}.

\section{Data collection and preprocessing} \label{sec:data}

Since February 2020, amazing efforts have been done to collect and share data at different levels. Many governments, public health institutions or media provide open data for tracking cases, number of deaths, etc. \citet{alamo2020open} propose an interesting and quite complete survey on the main open-resources for addressing the Covid- 19 pandemic from a data science point of view. In this paper, we use six different sources of time series data recorded daily. The data are slightly preprocessed in order to not pay too much attention on very small numbers of confirmed cases or deaths for small cities, provinces, states or regions. Datasets have been imported and preprocessed using the \texttt{R} software, which is also used to produce simulations and numerical results presented in the next section. Let us detail the data sources and the preprocessing step.

\begin{itemize}
\item Chinese data: we use the \texttt{R} package \texttt{nCov19} written by~\citet{wu2020open} which collects data at city level in China. Hubei province concentrates more than 97\% of reported deaths in China. We therefore consider the numbers of Hubei province and aggregate numbers of all others provinces. Inside Hubei province, we keep numbers for the cities of Wuhan, Xiaogan and Huanggang (92\% of reported deaths in Hubei province as of \study) . Other cities are not considered as they contained several artefacts. The data at the national level are finally added to the dataset. 
\item Canadian data: the Public Health Agency of Canada provides a daily report, available as a \texttt{csv} file\footnote{\url{https://www.canada.ca/en/public-health/services/diseases/2019-novel-coronavirus-infection.html}}, of confirmed cases and deaths for Canadian provinces. In addition to these data, we have also access to regional data from Quebec. Regional data compiled into a \texttt{json} file, were kindly provided by Antoine Béland web developer at \texttt{Le Devoir}\footnote{\url{https://www.ledevoir.com/documents/special/2020-03-25-tableau-de-bord-coronavirus/index.html}}. We aggregate the numbers of confirmed cases and deaths of provinces of Canada (resp. regions of Quebec) which have, as of \study, a cumulative number of deaths smaller than the first decile of the cumulative number of deaths of  Canada. We also add  data at the national level to the dataset.
\item US data: several sources of data can be found on the web. We use data reported by the COVID Tracking Project \url{https://covidtracking.com/} (an open data repository which is cited by several US newspapers)\footnote{\url{https://covidtracking.com/api/data}}. As we did for Canada, we aggregate (in the same way) states which have small numbers of cumulative numbers of deaths as of \study. Finally, national numbers reported by ECDPC (see below) are added to the dataset.
\item French data: the Public France Health System provides open data at the official open data portal \url{https://www.data.gouv.fr/}. We consider the regional data available as a \texttt{csv} file\footnote{\url{https://www.data.gouv.fr/fr/datasets/donnees-hospitalieres-relatives-a-lepidemie-de-Covid-19/}}. This dataset describes numbers of hospitalized cases and deaths at hospitals. As we did for Canada and the US, we aggregate small regions of France and data at the national level reported by ECDPC (see below) are added to the dataset.
\item International dataset: the European Centre for Disease Prevention and Control provides data at the national level available worldwide as a \texttt{csv} file\footnote{\url{https://www.ecdc.europa.eu/en/publications-data/download-todays-data-geographic-distribution-Covid-19-cases-worldwide}}. We use these data for France and the USA which seem to be more complete. For France for instance, the numbers of deaths take into account the ones which appear in nursing homes. 
\end{itemize}

All Data are recorded daily. They are collected since December 2019 for China, February 1 2020 for Canada, February 29 2020 for the US,  March 18 for France and December 31, for the international dataset. For all datasets except the French one, we have at our disposal cumulative and daily numbers of confirmed cases and deaths. For France, the numbers of confirmed cases are not available and are substituted by numbers of hospitalized patients. In the rest of this paper, we make an abuse and speak of confirmed cases for hospitalized patients in France. Depending on the national public policies, the number of reported cases and deaths are not standardized, which has given rise to massive and heated debates. It is not the intention of this paper to present or discuss these policies. We refer the reader to the different websites to understand how a case or death is considered for the different countries (note that even between two US states or Canadian provinces, the way numbers are reported are different).

The study period goes from December 1, 2019 to \study\, when we analyze daily numbers of confirmed cases or deaths. When we analyze cumulative data, we have to be careful as some  cities, provinces or countries have already passed the epidemiological peak, meaning that the cumulative number of cases or deaths remain almost unchanged several days or weeks in a row. For the city of Wuhan for instance, the cumulative number of deaths has remained between 2000 and 3000 since February, 25 and for several weeks in a row. So there would be a clear overexpression of the digit 2 for this time series data if we were to keep the same study period as for analyzing daily data. Thus, when we analyze cumulative data,  we stopped the study period as of February 15, 2020 for Chinese data and April, 15 for other ones. Keeping cumulative numbers in the data analysis is highly relevant as there is a consensus in the epidemiological literature that  numbers of cases or deaths grow exponentially and exponential curves are known to follow quite well the Newcomb-Benford distribution \citep[see e.g.][]{berger2015introduction}.

Tables~\ref{tablen} summarizes this section and provides as of \study, the sample sizes available for analyzing first or second digit for each dataset. Obviously, for the second digit, we focus on numbers of confirmed cases and deaths that are larger than 10.

\begin{table}[ht]
\centering
\begin{tabular}{rcccccccc}
  \hline 
 \rowcolor{gray!20} &  \multicolumn{4}{c}{Daily data} &\multicolumn{4}{c}{Cumulative data} \\
 \rowcolor{gray!20}& \multicolumn{2}{c}{1st digit} &\multicolumn{2}{c}{2nd digit} 
                          & \multicolumn{2}{c}{1st digit} &\multicolumn{2}{c}{2nd digit}  \\
 \rowcolor{gray!20}& Cases & Deaths &Cases & Deaths  & Cases & Deaths&Cases & Deaths    \\
 \hline
China & 501 & 447 & 359 & 155 & 282 & 213 & 261 & 154 \\ 
  Canada & 631 & 362 & 522 & 167 & 394 & 163 & 341 & 135 \\ 
  USA & 1658 & 1326 & 1484 & 902 & 1071 & 808 & 985 & 637 \\ 
  France & 579 & 562 & 565 & 406 & 336 & 328 & 332 & 299 \\ 
   \hline
\end{tabular}
\caption{Sample size available for analyzing first or second digits  of daily  and cumulative numbers of confirmed cases (Cases) and deaths (Deaths) for different countries. The study period goes from December 2019 to \study \; for daily data and from December 2019 to mid-February (resp. mid-April) for Chinese data (resp. other data).} 
\label{tablen}
\end{table}

\section{Results} \label{sec:results}

\begin{figure}[ht]
\begin{center}
\includegraphics[scale=.66]{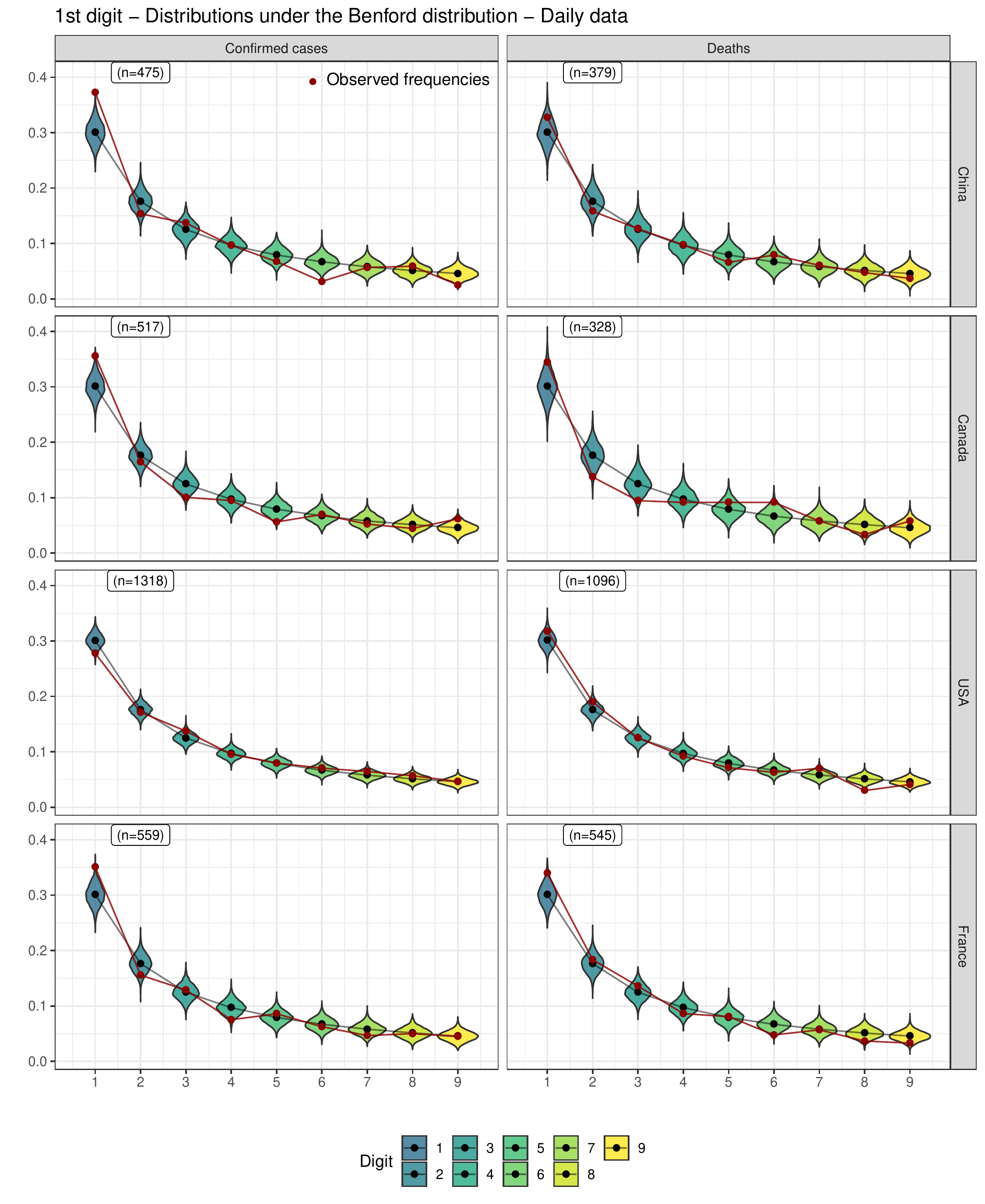}	
\caption{Frequencies of first digits from {\bf daily numbers} of confirmed cases and reported deaths for different countries. Black points and curve correspond to the Newcomb-Benford probabilities for the first digit. Violin boxplots are Monte Carlo estimates of the distribution of proportions of the first digit under the Newcomb-Benford distribution. Boxplot are constructed using $B=\BB$ simulations with sample size $n$.}
\label{boxplot1}
\end{center}
\end{figure}

\begin{figure}[h]
\begin{center}
\includegraphics[scale=.66]{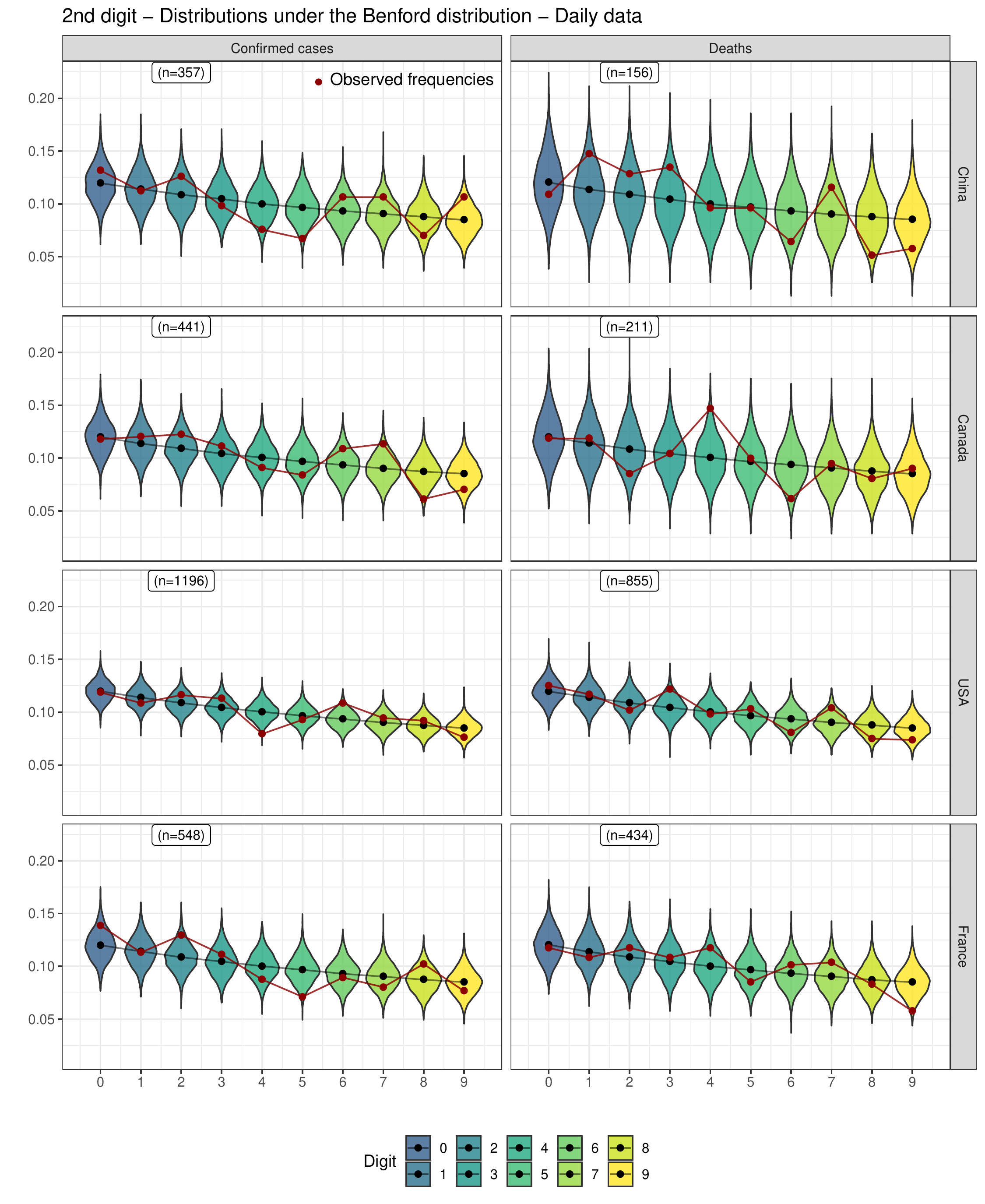}	
\caption{Frequencies of second digits from {\bf daily numbers} of confirmed cases and reported deaths for different countries. Black points and curve correspond to the Newcomb-Benford probabilities for the second digit. Violin boxplots are Monte Carlo estimates of the distribution of proportions of the second digit under the Newcomb-Benford distribution. Boxplot are constructed using $B=\BB$ simulations with sample size $n$.}
\label{boxplot2}
\end{center}
\end{figure}

\begin{figure}[ht]
\begin{center}
\includegraphics[scale=.66]{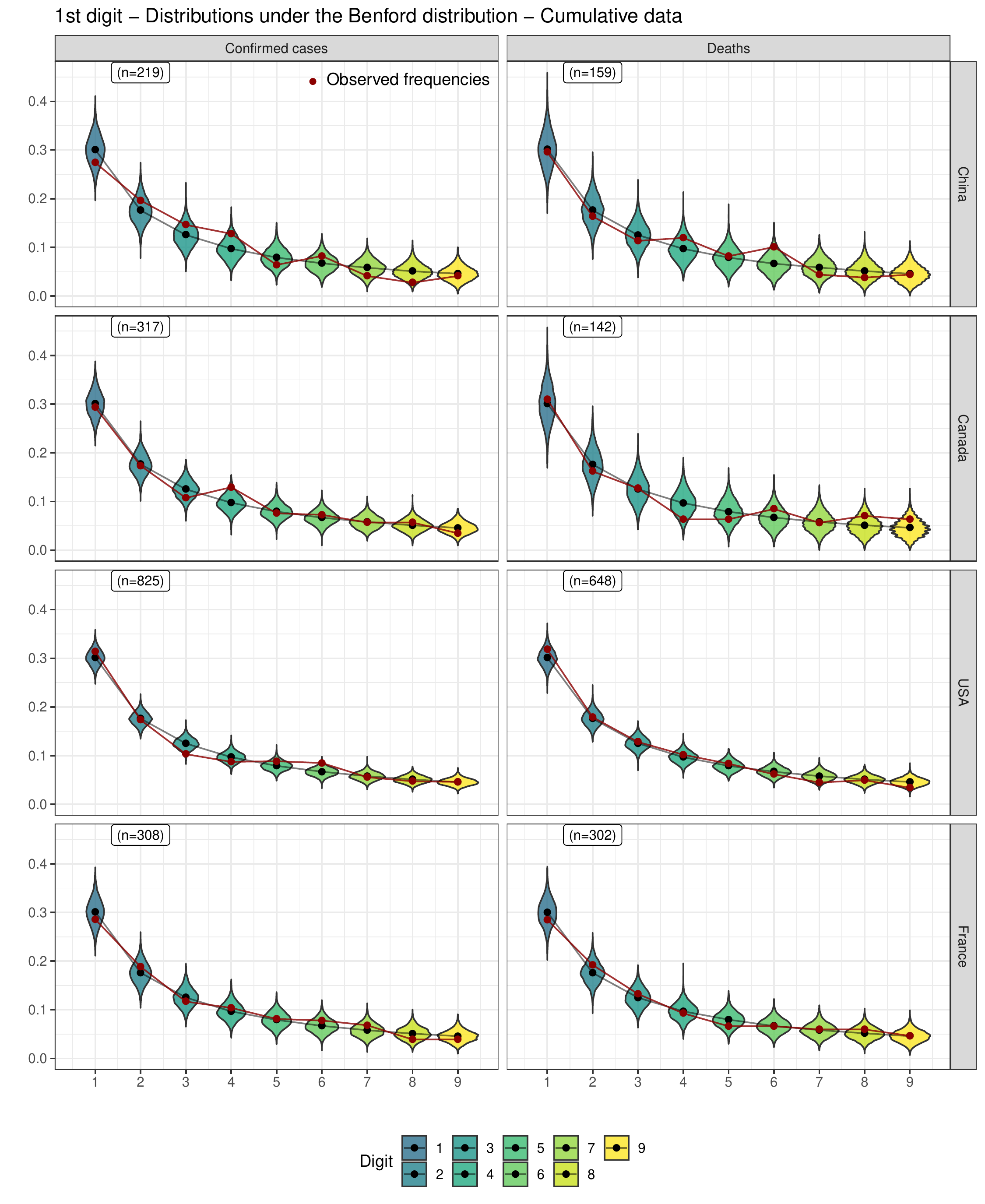}	
\caption{Frequencies of first digits from {\bf cumulative numbers} of confirmed cases and reported deaths for different countries. Black points and curve correspond to the Newcomb-Benford probabilities for the first digit. Violin boxplots are Monte Carlo estimates of the distribution of proportions of the first digit under the Newcomb-Benford distribution. Boxplot are constructed using $B=\BB$ simulations with sample size $n$.}
\label{boxplot3}
\end{center}
\end{figure}

\begin{figure}[h]
\begin{center}
\includegraphics[scale=.66]{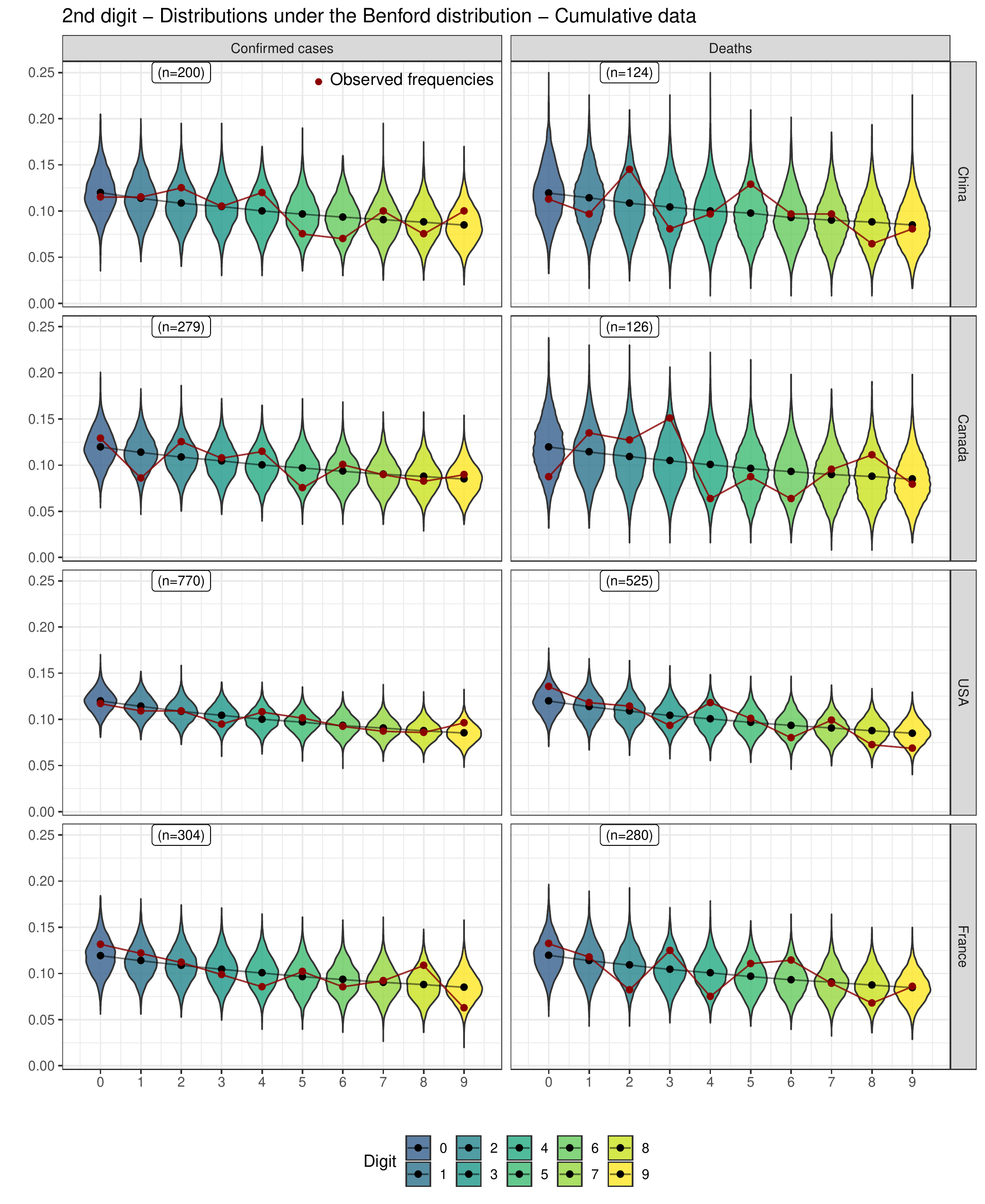}	
\caption{Frequencies of second digits from {\bf cumulative numbers} of confirmed cases and reported deaths for different countries. Black points and curve correspond to the Newcomb-Benford probabilities for the second digit. Violin boxplots are Monte Carlo estimates of the distribution of proportions of the second digit under the Newcomb-Benford distribution. Boxplot are constructed using $B=\BB$ simulations with sample size $n$.}
\label{boxplot4}
\end{center}
\end{figure}

Figures~\ref{boxplot1}-\ref{boxplot4} summarize the efforts of this data analysis. For daily numbers (Figures~\ref{boxplot1}-\ref{boxplot2}) and cumulative numbers (Figures~\ref{boxplot3}-\ref{boxplot4}) of confirmed cases and deaths reported for China, Canada, USA and France, we estimate proportions of digits 1,2,\dots,9 (for the first digit) and 0,1,\dots,9 (for the second one). This information is represented by red curves (observed frequencies) in Figures~\ref{boxplot1}-\ref{boxplot4}. For each dataset (i.e. for first  or second digit analysis, each type of data and each country), we also measure empirically departures from the Newcomb-Benford distribution. Thus, boxplots correspond to estimates of proportions of first or second digit based on $B=\BB$ simulations of sample size $n$ from the Newcomb-Benford distribution. These figures do not constitute a formal goodness-of-fit test. This will be examined later taking into account the multiplicity of tests. However, it is clear that the Newcomb-Benford distribution seems to be an excellent candidate.

Small departures to the Newcomb-Benford distribution seem noticeable in Figures~\ref{boxplot1} and \ref{boxplot3}: for instance digits 1 and 6 (resp. 1) for daily numbers of confirmed cases in China (resp. China), digit 3 for daily numbers of deaths in Canada, etc. Small departures are also noticeable in Figures~\ref{boxplot2} and \ref{boxplot4}. Because the sample size slightly decreases when analyzing second digits, violin boxplots exhibit slightly higher dispersion. 

The first and clear conclusion drawn from Figures~\ref{boxplot1}-\ref{boxplot4} is that daily and cumulative numbers of confirmed cased and deaths exhibit a first digit phenomenon. Digit 1 is the most frequent, followed by 2, etc. As a second general conclusion, it does appear that the empirical distributions of digits for one type of data seem quite similar across countries. Overall, we also remark that the distribution of digits seem to be closer to the Newcomb-Benford distribution for cumulative data than for daily data. 

\begin{figure}[h]
\hspace*{-1.5cm} \includegraphics[scale=.78]{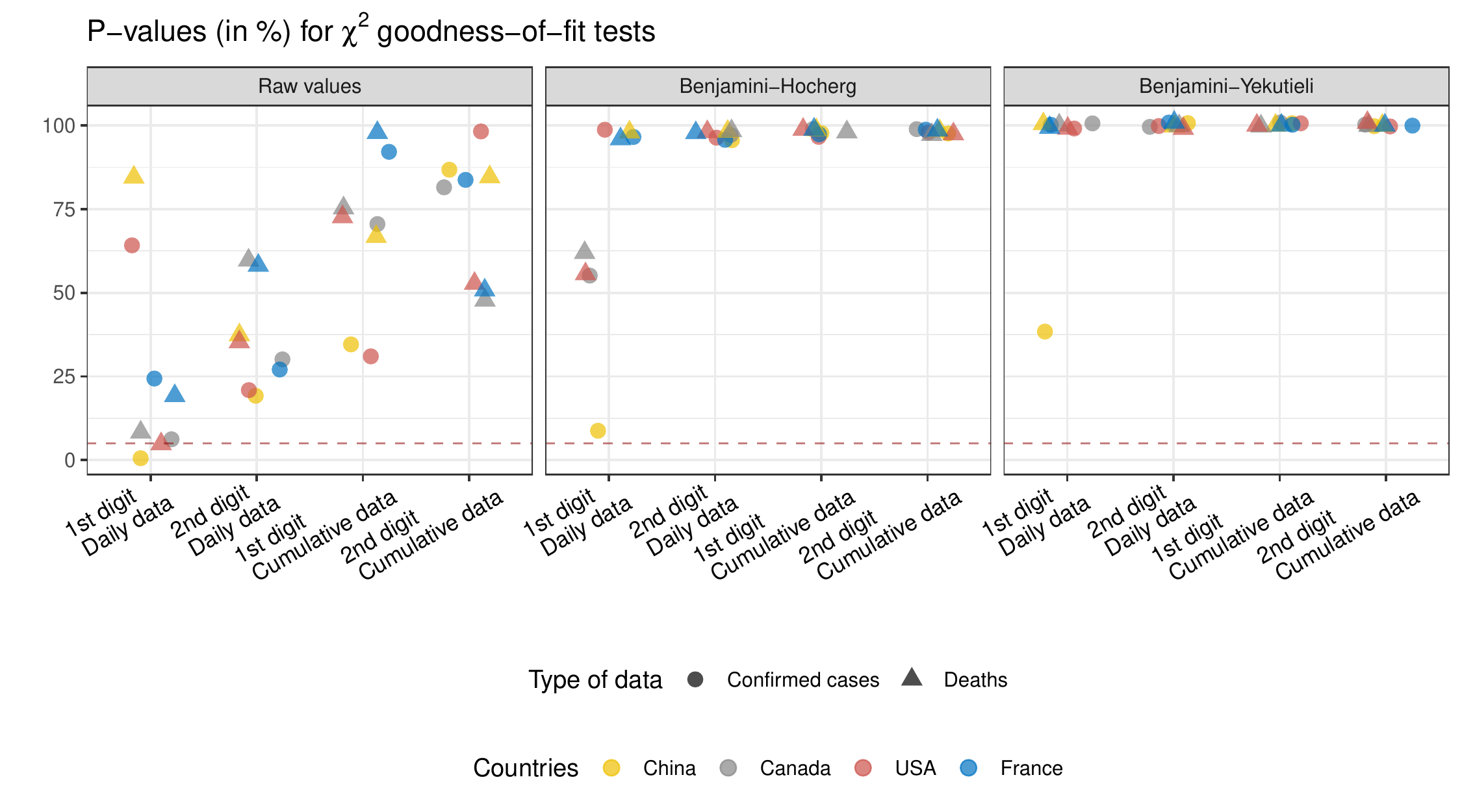}	
\caption{P-values (in \%) of the $\chi^2$ goodness-of-fit tests of the Newcomb-Benford distribution for the first  and second digits for the reported daily and cumulative confirmed cases and deaths for different countries. P-values are obtained via Monte Carlo using $B=\BB$ replications. The left column corresponds to raw values, the middle one to values adjusted using the Benjamini-Hochberg's procedure, while the right column to adjusted p-values  using the Benjamini-Yekutieli's procedure. Dashed red line corresponds to the level 5\%. Points are slightly jittered for a better visualization.}
\label{pvals}
\end{figure}

To continue this data analysis, Figure~\ref{pvals} provides a more formal and quantitative approach. For each of the 32 datasets (2 digits, 4 countries, confirmed cases/deaths, daily/cumulative data), we perform a $\chi^2$ goodness-of-fit test to judge the adequacy of the Newcomb-Benford distribution. Since the sample size is not large for some datasets, we estimate the distribution of the standard $\chi^2$ statistic under $H_0$ using Monte Carlo approach ($B=\BB$ replications are used). Figure~\ref{pvals} reports adjusted p-values in percentage. Given the quite large number of tests done in this paper, all p-values in this manuscript are adjusted as a whole, using a false discovery control procedure. Procedures which control false discovery rate are subject to assumptions on the distribution of p-values. The most well-known procedure is Benjamini-Hochberg's  (BH) procedure \citep{benjamini1995controlling} which requires a PRDS assumption (positive regression type dependency, see \citet{benjamini2001control}). This assumption seems really complex and is probably wrong in the context of this paper: the number of deaths depends on the number of confirmed cases, the distribution of first digit is not independent of the distribution of second digit and cumulative data depend on daily data. All these characteristics have influence on the dependence on p-values. Thus, we also adjust p-values using  the Benjamini-Yekutieli's (BY) procedure \citep{benjamini2001control} which allows an FDR control under any type of dependence. Although BY's procedure is more conservative than the BH's procedure, it does control theoretically FDR at level 5\%. Figure~\ref{pvals} presents raw values and adjusted p-values using BH and BY's procedures. Figure~\ref{pvals} reveals that no discovery can be made at FDR level 5\%. The smallest BY's adjusted p-value equals 38.5\% (note that even the smallest BH's adjusted p-value, 8.6\%, is larger than 5\%) and most of adjusted p-values are close or equal to 100\%. Clearly, as every (omnibus) goodness-of-fit test should be interpreted, this does not prove that each dataset should be modeled by a Newcomb-Benford distribution (for the first or second digit) but tends to convince us that these models are legitimate. 

\begin{figure}[h]
\begin{center}
\includegraphics[scale=.78]{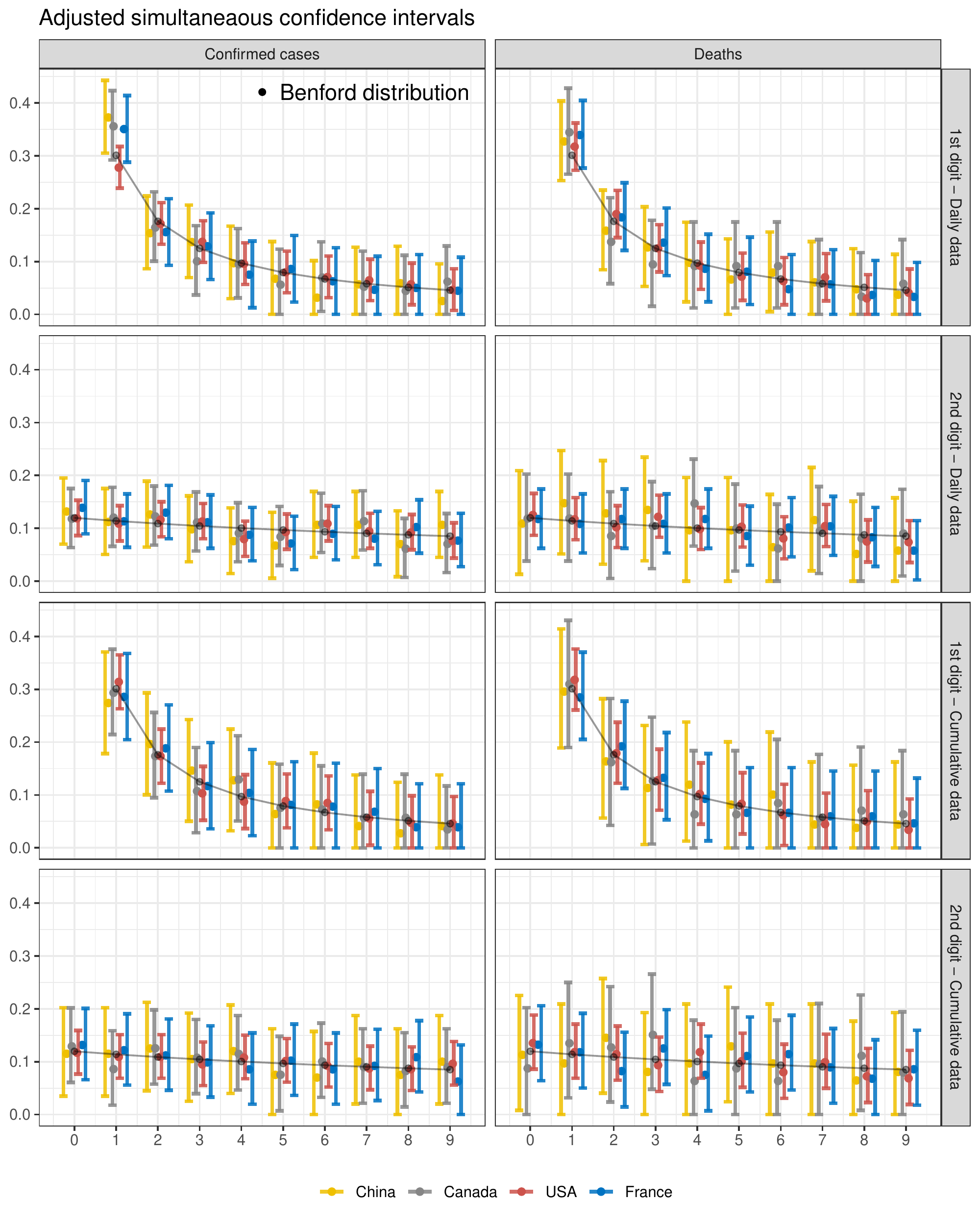}	
\caption{Simultaneous 95\% confidence intervals for first  and second digits proportions based on reported daily and cumulative numbers of confirmed cases and deaths  for different countries. 
Black points and curve correspond the probabilities of first and second digits under the Newcomb-Benford distribution.}
\label{ci}
\end{center}
\end{figure}

\begin{figure}[h]
\hspace*{-1.5cm} \includegraphics[scale=.78]{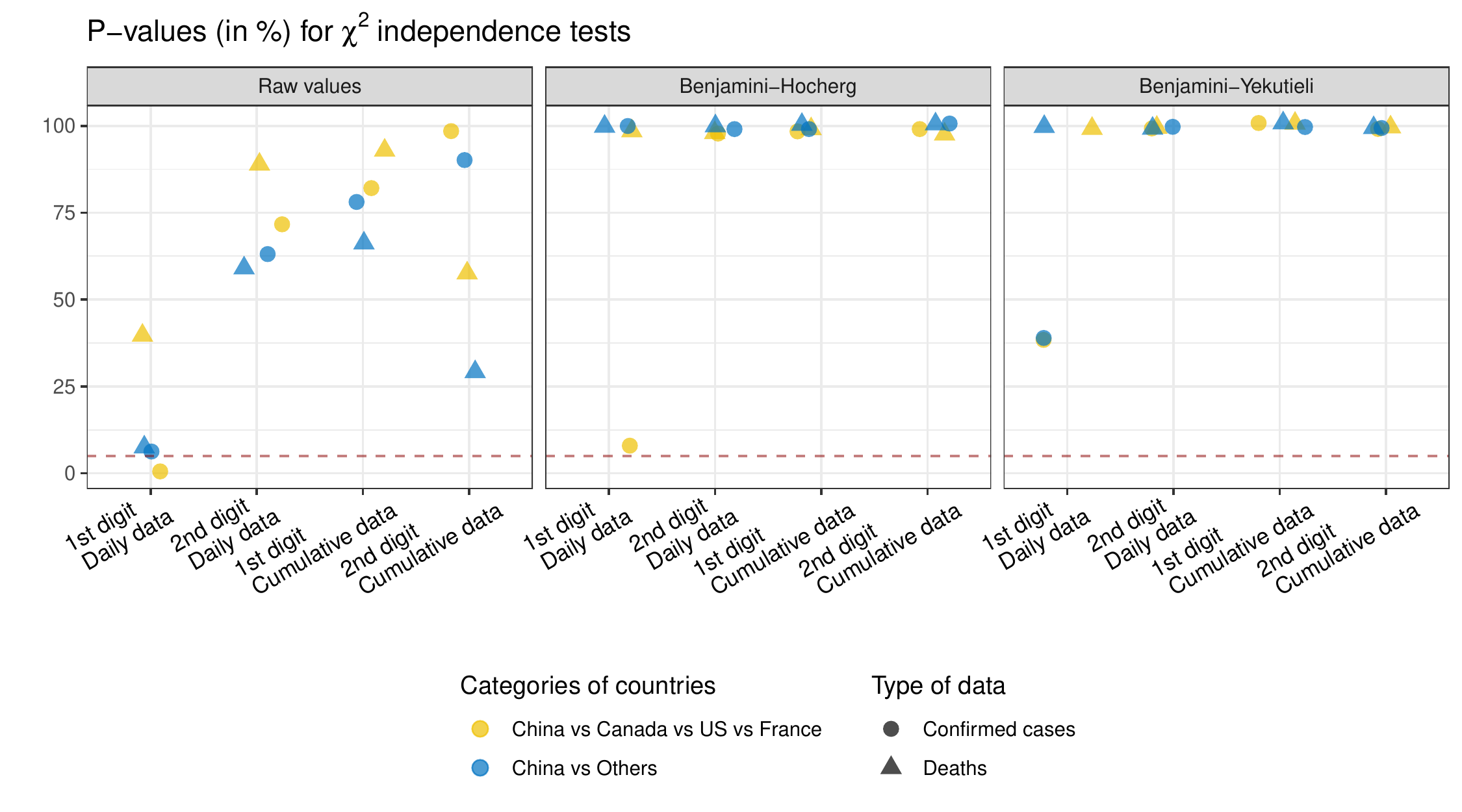}	
\caption{P-values (in \%) of the $\chi^2$ independence goodness-of-fit tests between digits (first  and second digits for daily and cumulative numbers of confirmed cases and deaths) and the variable ``Countries''. The latter is either composed of the four countries, or two countries China and the other ones which are aggregated. P-values are obtained via Monte Carlo using $B=\BB$ replications. The left column corresponds to raw values, the middle one to values adjusted using the Benjamini-Hochberg's procedure, while the right column to adjusted p-values using the Benjamini-Yekutieli's procedure. Dashed red line corresponds to the level 5\%. Points are slightly jittered for a better visualization.}
\label{pvalsGroup}
\end{figure}

Figure~\ref{ci} is less focused on the Newcomb-Benford model. We aim at measuring differences between the datasets by constructing 95\% simultaneous confidence intervals. Different solutions for the computation of simultaneous confidence intervals for a single multinomial distribution have been proposed in the literature. We consider, here, the approximation proposed by~\citet{sison1995simultaneous}, implemented in the \texttt{R} package \texttt{MultinomialCI}. 
To take into account the multiplicity of confidence intervals (32 intervals of multinomial distributions in total), we simply apply the Bonferroni correction. We also report the Newcomb-Benford probabilities as a reference. The conclusion follows along the same lines as previous figures. It seems impossible to draw any firm conclusion from Figure~\ref{ci} that one dataset behaves  differenty from other ones. Figure~\ref{pvalsGroup} strenghtens this comment. We report p-values for $\chi^2$ independence tests, where we test the distribution of counts for one dataset against the variable country. By count dataset, we mean here the distribution of counts for first  or second digit, for daily or cumulative numbers of confirmed cases or deaths. For each such count distribution, we  analyze through $\chi^2$ independence tests, either differences between different countries, or difference between China and other countries. All p-values (estimated using a Monte Carlo approach with $B=\BB$ replications) are adjusted using BH's procedure as well as BY's procedure. Again, Figure~\ref{pvalsGroup} shows that no null hypothesis can be rejected at FDR level 5\% since the smallest BY's adjusted is much larger than 5\%. Thus, there is no clear evidence of any difference between the different countries considered in this study or between China and others.

\section{Discussion} \label{sec:discussion}

The Newcomb-Benford distribution is often used on real datasets to detect potential wrong reports of numbers. Applied to Covid-19 daily and cumulative numbers of confirmed cases and deaths for China, Canada, USA and France, it is  clear that these data exhibit the first digit phenomenon as initially observed by \citet{newcomb1881note}. We find out that the Newcomb-Benford distribution for the first and second digits cannot be rejected at a false discovery rate 5\%.  This does not prove that the leading digits of epidemiological data should be modeled by Newcomb-Benford distribution. Nevertheless, it opens an interesting research question: could we prove that the first digits of data that fit SIR models (or extensions) follow the Newcomb-Benford distribution? This is clearly out of the scope of this note.

Putting aside the Newcomb-Benford model, our analysis shows that there are no qualitative differences between data from the Chinese government and from other countries (considered in this study), although that sources of biases for all datasets are very large (for instance the province of Quebec stands out for reporting also deaths for suspected people to Covid-19 which was not the case in France before the beginning of April).

This data analysis has its obvious limitations. Showing that frequencies of digits are close to expected probabilities, or showing that frequencies between two countries are quite similar does neither prove nor disprove that there was a fraud in the reported numbers. If, for instance, one multiplies by 10 each daily number of deaths or confirmed cases, the results remain unchanged! However, as mentioned in the introduction, Newcomb-Benford distribution has been applied to many real-life datasets and it is interesting to see, as perhaps might have been expected, that this model also shows up when analyzing Covid-19 data.

\section*{Acknowledgements}

The author is grateful to Frédéric Lavancier, Christian Genest and to colleagues from the Department of mathematics at UQAM and in particular Genevi\`eve Lefebvre, Sorana Froda, Arthur Charpentier for their careful reading, suggestions and comments. The author would also like to thank Antoine B\'eland for providing access to data in the province of Quebec.

%%%%%%%%%%%%%%%%%%%%%%%%%%%%%%%%%%%%%%%%%%%%%%%%%%%%%%%%%%%%%%%%%%%%%%%%%%%
%%%%%%%%%%%%%%%%%%%%%%%%%%%  End of manuscript text
%%%%%%%%%%%%%%%%%%%%%%%%%%%%%%%%%%%%%%%%%%%%%%%%%%%%%%%%%%%%%%%%%%%%%%%%%%%

\section*{Supporting information}

The following supporting information is available as a \texttt{zip} on the webpage of the author\footnote{\url{https://sites.google.com/site/homepagejfc/publications}}. The file \texttt{covidBenford.zip} contains:
\begin{itemize}
	\item \texttt{data.R}: \texttt{R} code file used to import data from online resources and used to preprocess the data as  described in Section~\ref{sec:data}.
	\item \texttt{allData2020-05-11.RData}: \texttt{Rdata} file corresponding to data as of \study. This file contains the dataframes: \texttt{df.glob}, \texttt{df.hubei}, \texttt{df.can}, \texttt{df.usa} and \texttt{df.france}.
	\item \texttt{covidBenford.R}: \texttt{R} code used to prepare Figures in this manuscript.
	\item \texttt{covidBenford.Rmd}: knitting this \texttt{Rmarkdown} file allows the reader to reproduce figures of this manuscript. 
\end{itemize} 
Note that this study is dynamic as data are still being collected. The  \texttt{Rmarkdown} file allows the reader to run the code based on updated data. However, the construction of the database depends on web resources. Therefore, from \study, the author is not responsible of possible errors that would appear due to a change in the \texttt{R} package \texttt{nCov2019}, or the \texttt{csv} and \texttt{json} files described in Section~\ref{sec:data}.

\bibliographystyle{plainnat}
\bibliography{benford}

%%%%%%%%%%%%%%%%%%%%%%%%%%%%%%%%%%%%%%%%%%%%%%%%%%
%%%%%%%%%%%% Boxplots 
%%%%%%%%%%%%%%%%%%%%%%%%%%%%%%%%%%%%%%%%%%%%%%%%%%

%%%%%%%%%%%%%%%%%%%%%%%%%%%%%%%%%%%%%%%%%%%%%%%%%%
%%%%%%%%%%%% pvaleurs
%%%%%%%%%%%%%%%%%%%%%%%%%%%%%%%%%%%%%%%%%%%%%%%%%%

%%%%%%%%%%%%%%%%%%%%%%%%%%%%%%%%%%%%%%%%%%%%%%%%%%
%%%%%%%%%%%% simultaneous confidence intervals
%%%%%%%%%%%%%%%%%%%%%%%%%%%%%%%%%%%%%%%%%%%%%%%%%%

%%%%%%%%%%%%%%%%%%%%%%%%%%%%%%%%%%%%%%%%%%%%%%%%%%
%%%%%%%%%%%% pvaleurs groupe
%%%%%%%%%%%%%%%%%%%%%%%%%%%%%%%%%%%%%%%%%%%%%%%%%%

\end{document}